\documentclass[aps,eqsecnum,showpacs,twocolumn]{revtex4}

\usepackage{graphicx}

\begin{document}
\title{Chaotic quantum ratchets and filters with cold atoms in optical 
lattices: properties of Floquet states}
\author{G.~Hur, C.E.~Creffield, P.H.~Jones and T.S.~Monteiro}
\affiliation{Department of Physics and Astronomy, University College London,
Gower Street, London WC1E 6BT, UK}
\date{\today}

\begin{abstract}
Recently, cesium atoms in optical lattices subjected to cycles
of unequally-spaced pulses have been found to show 
interesting behavior: they represent the first experimental
demonstration of a Hamiltonian ratchet mechanism, and  
they show strong variability of the Dynamical Localization lengths
as a function of initial momentum.
The behavior differs qualitatively from corresponding atomic systems pulsed with
equal periods, which are a textbook implementation of a well-studied
quantum chaos paradigm, the quantum $\delta$-kicked particle ($\delta$-QKP).
We investigate here the properties of the
corresponding eigenstates (Floquet states) in the parameter regime
of the new experiments and compare them with those of the
eigenstates of the $\delta$-QKP at similar kicking strengths.
We show that, with the properties of the Floquet states, we can 
shed light on the form
of the observed ratchet current as well as variations in the
Dynamical Localization length. 
\end{abstract}

\maketitle

\section{Introduction}
Periodically-kicked quantum systems, such as the
$\delta$-kicked particle ($\delta$-KP), 
have long played a central role in studies of quantum chaos and the
correspondence between quantum behavior and the underlying
classical dynamics \cite{Casati,Fish}.
Advances in the manipulation of
cold atoms have permitted the experimental realization of these
systems in pulsed optical lattices \cite{Raizen}. Experiments  with
sodium and cesium atoms have demonstrated
the phenomenon of ``Dynamical Localization'' (DL) --
the quantum suppression of classical chaotic diffusion --
and established the suitability of these systems
as an arena for the study of effects arising from
quantum chaos.

When treating conservative quantum systems it is 
frequently useful to study the system's energy-eigenstates, and
for periodically driven systems the appropriate generalization of 
these states is given by the Floquet states.
This approach has provided extensive insight into the properties of 
the standard QKP,
and has shown, for example, that DL arises directly
from the exponential localization of the system's Floquet
states \cite{Fish}. Observed momentum
oscillations associated with
 chaos-assisted tunneling, in experiments using periodically-driven 
cold atoms \cite{Raizcat} and
BECs \cite{Phillips} have also been 
analysed with Floquet theory; it was found that
the oscillation period is determined by the  splittings of the Floquet phases
of a pair of symmetry-related eigenstates. 
The statistics of QKP Floquet quasi-energy spectrum have been
studied extensively see e.g. \cite{Izrael} and compared
with the predictions of Random Matrix Theory.
Notably, though, the $\delta$-QKP has Poissonian
short-range statistics (which are typically associated with integrable dynamics) 
even for very large values of $K$, where the dynamics is fully chaotic. This has
been shown to be a further consequence of DL \cite{Izrael}.

However, a series of recent theoretical 
\cite{Mont,Jonck} and experimental \cite{Jones,Jon1,PJones} studies of
cold atom systems subjected to repeating cycles of unequally
spaced kicks revealed dynamics rather different from that found in
the corresponding standard QKP systems. Two types of unequally 
$\delta$-kicked systems were investigated. The first \cite{Jonck,Jones} 
consists of a $\delta$-KP with small
perturbations made to the kick-period. We term it the perturbed-period KP. 
In the second system, \cite{PJones} the system is periodically
subjected to {\em pairs} of closely-spaced kicks. This is referred to
as  the double $\delta$-KP  or 2$\delta$-KP. 

In a theoretical study, the perturbed-period KP was found to yield a 
quantum momentum current even in the chaotic regime  
\cite{Mont,Jonck}.
This was unexpected in a chaotic Hamiltonian system, since
to date only mixed phase-space ratchet mechanisms had been investigated
\cite{Ditt,Flach}. A simple definition of a ratchet 
is a spatially periodic device
which produces a current without net bias (ie the time and space-averaged
forces are zero). Most of the extensive ratchet literature deals with
dissipative or Brownian ratchets \cite{Reimann} and comparatively 
little theoretical 
work has been undertaken on Hamiltonian ratchets, which 
are dissipation and noise-free.
In  \cite{Jones,Jon1},  a momentum distribution with a non-zero
average (constant in time) was obtained experimentally from an atomic 
cloud with initial zero average momentum. We are
unaware of any other experimental studies of Hamiltonian quantum ratchets:
all implementations to date have been of
dissipative/Brownian ratchets. Hence the results from \cite{Jon1} 
and also reproduced here, represent
the only implementation of a type of Hamiltonian quantum ratchet,
whether chaotic or mixed phase-space.
In Ref. \cite{Jonck} it was also proposed that the chaotic diffusive properties 
of the perturbed-period KP could be exploited to filter cold
atoms, according to their momenta, by controlling the Dynamical Localization.

For the second system, the 2$\delta$-KP, a theoretical and experimental
study  \cite{PJones}
revealed that the diffusion is dominated by 
long-ranged correlations which control escape from well-defined
momentum trapping regions. This combination of strong chaotic
diffusion and long-ranged correlations is novel: strong
chaos is generally associated with rapidly decaying correlations.

It is clear that Floquet theory is central to the analysis of 
chaotic, time-periodic quantum systems. The need to understand further
the chaotic Hamiltonian ratchet as well as the 2$\delta$-KP motivated
this study of the Floquet states of these systems.
The paper is organized as follows. 
In the next section we review the well-known $\delta$-KP, then introduce the
perturbed $\delta$-KP and the double $\delta$-KP systems. 
In Section III we give a brief review of the Floquet approach.
In Section IV we compare the results with recently obtained experiments
on these systems.
Finally in Section V we give our conclusions.

\section{Introduction to $\delta$-kicked systems}

The Hamiltonian for the usual $\delta$-KP can be written as
\begin{equation}
\label{ham_kp}
H = \frac{p^{2}}{2}+ K \sin x \sum_n \delta(t-nT)
\end{equation}
where $K$ is the kick strength and $T$ is the time interval
between successive kicks. Consider its effect on  an ensemble of
particles with a gaussian momentum distribution centered
on $p_0$, $N(p)= \exp[-(p-p_0)^2/\Delta p^2]$.
The classical dynamics depends only on the parameter $K$,
and for values larger than $K  \simeq 1$, 
the chaotic diffusion is not bounded by classical barriers.
In this regime the ensemble will diffuse in momentum space,
its average energy growing linearly with time as 
$\langle p^2 \rangle = D t$, 
where, to lowest order, the diffusion rate is given by
$D_0 = K^2/2$. The distribution will thus remain 
gaussian, although its width will increase with time as $\Delta p(t)=
\sqrt{Dt}$. 

In contrast, the quantum system only follows
this behavior up to a timescale $t^{\ast} \simeq D/ \hbar^2$
\cite{Shep}, after which diffusion in momentum space is
suppressed -- dynamical localization (DL). Such a system
will asymptotically evolve towards a characteristic exponential
momentum distribution, $N(p) \sim \exp \left[ -|p-p_0|/\Delta p_Q \right]$, 
with constant width $\Delta p_Q \sim \sqrt{Dt^*} \sim D/\hbar$,
which thus acts as an experimental fingerprint for DL \cite{Raizen}.
As DL is a wave-coherent effect, the
quantum system must preserve coherence over at least
the timescale $t^{\ast}$
for this effect to be observable. Indeed, it has been
verified experimentally \cite{Raizen2} that the DL profile does
not survive the presence of noise or dissipation, and that with
decoherence a more gaussian profile for $N(p)$ will be produced.

In both classical and quantum cases the behavior of the 
standard $\delta$-KP is essentially
independent of $p_0$ since, even for modest values of $K$, the effects of
small fluctuations in the structure of phase-space
are on negligible scales relative to $\Delta p_Q$. Even if there are
small stable islands, they are of size $\Delta p \sim 1 $ so have
little effect on the general form of $N(p)$, since typically
$\Delta p_Q \gg 1$. 

The classical dynamics of the $\delta$-KP is obtained by iterating
the well-known ``Standard Map''. For
 the perturbed-period and double $\delta$- kicked 
systems, on the other hand, the dynamics is given by a 2-kick map:
\begin{eqnarray*}
p_j &=& p_{j-1} - V'(x_{j-1})  \\
x_j &=& x_{j-1} + p_{j} T_1 \\
p_{j+1} &=& p_{j} - V'(x_j)  \\
x_{j+1} &=& x_j + p_{j+1} T_2 . \
\end{eqnarray*}
Clearly, setting $T_1=T_2$ and $V'(x)=K \cos x$, we recover the Standard Map.
For the perturbed-period KP, the lengths of
the two kicking periods are $T_1= 1+\epsilon$ and
$T_2=1- \epsilon$, where $\epsilon \ll 1$. The
perturbation thus consists of slightly
altering the kicking period about its mean.
For the double $\delta$-KP we take 
$T_1= 2 - \epsilon$, $T_2=\epsilon$, although we shall also
show the effect of interchanging the two kick periods.
It should be noted that these systems
are {\em time-periodic}, with period $T_{tot}=T_1+T_2$, 
and are thus quite distinct
from the recent interesting study of two independent kicking
sequences, which can be non-periodic and hence non-localizing
\cite{Garreau}.

As in the standard map, we consider a sinusoidal potential $V(x) = K \sin x$.
However, to obtain a ratchet current in the case of the perturbed-period KP,
we need to break the spatio-temporal symmetries, and  so we add an additional
``rocking'' linear potential of strength $A$. In this case the form
of the potential is $V(x) = -[K \sin x+Ax(-1)^j]$, where $j$ is
the kick number. In experimental implementations of this system, the rocking
linear term was obtained by means of an accelerated lattice \cite{Jones}.

We first consider in general terms how the introduction of the second timescale
$\epsilon$ modifies the classical behavior of the
standard map.  If we neglect all correlations, the standard
map has a constant momentum diffusion rate,
$D_0 \simeq K^2/2$  -- this is what one would expect if the 
momenta at consecutive kicks are uncorrelated and
so evolve as a random walk. However, unless $K$ is
exceedingly large, the time-evolution of the standard map will 
contain some short-range (2-kick and 3-kick) correlations. 
Including these corrections yields a modified diffusion constant
$D = \frac{K^2}{2}[1 - 2J_2(K) - (J_1(K))^2 \dots]$. Of
particular interest is the $J_2(K){K^2}$ term, representing
correlations $\langle V'(x_j) V'(x_{j+2}) \rangle$ between 
nearest-but-one kicks (the 2-kick correlation). 

For the modified systems it is also possible to obtain 
analytically the important correlations \cite{Jonck,PJones}.
For instance, for the perturbed-period system the corrected diffusion is
$D \simeq\frac{K^2}{2}[1 - 2J_2(K)\cos(2p_0\epsilon-A) - (J_1(K))^2 \dots]$.
We see that we have a modified 2-kick correlation 
which  {\em oscillates} with the initial momentum $p_0$.
This effect is clearly most significant for values of $K$ 
such that $2J_2(K) \sim 1$.
 
The key point is that perturbing the kick spacings $T$ by a small 
amount can result in
large scale (relative to $\Delta p_Q \sim D/\hbar$) variations in the
classical momentum diffusion, and that  these are present even in fully
chaotic regimes 
(we take this to mean the absence of visible
stable structures on the Poincar\'e surface of section). 
For the analysis of experiments, one must now consider a {\em local} 
diffusion rate $D(p_0)$, which depends on the initial relative momentum 
between the atoms and the optical lattice.

In \cite{Jones}, the perturbed-period system was implemented 
experimentally with a cloud of cesium atoms for the case $A=0$. It was verified
that the energy absorbed by the cloud after Dynamical Localization
$\langle (p-p_0)^2 \rangle \propto \cos(2p_0\epsilon)$ as expected. 
However, $A=0$ 
corresponds to a symmetric potential.
The case $A=\pi/2$ in the perturbed-period system
is particularly interesting since then the momentum
diffusion is asymmetric about $p_0 = 0$. This implies that atoms
with positive momenta will absorb kinetic energy at different rates from
those with momenta of the same magnitude but moving in the
opposite direction.
 This asymmetric momentum diffusion represents a type of fully chaotic
{\em momentum} ratchet: in other words, roughly equal numbers of particles
will diffuse to the left or to the right, but those diffusing to the right,
on average, move faster.

 In a further experimental study of the perturbed-period KP \cite{Jon1},
the $A=\pi/2$ potential was implemented by means of an accelerated
lattice. It was found that an atomic cloud prepared
initially with a gaussian momentum distribution centered 
on $p_0=0$  evolved into a
distribution with non-zero, but constant, $\langle p \rangle$ 
which persisted even
beyond the break-time $t^{\ast}$, as expected from the theory \cite{Mont}.
This type of chaotic directed motion was first identified
in a  slightly different system: a 
kicked asymmetric double-well potential \cite{Mont}.
However, the latter potential gives rise to a rather
more complicated diffusive behavior, and also has proved much harder
to implement experimentally.  For these reasons here we do
not consider the case of the asymmetric double-well ratchet, but note
that our Floquet analysis of the perturbed-period KP can be carried-over to the
system investigated in \cite{Mont}.

The second system we consider explicitly in this work, the 2$\delta$-KP,
has diffusive behavior which is qualitatively
different to both the standard map and perturbed-period KP. While for
these other kicked systems we can analyze the diffusion as an
uncorrelated term, $K^2/2$, corrected by short-ranged correlations
(typically only 2 or 3-kick correlations for $K \simeq 3$), for the double 
$\delta$-KP, we find that the diffusion at long times is dominated by
families \cite{PJones} of long-ranged ``global'' correlations
(``global'' in the sense that they correlate all kicks up to the
time under consideration). At short times, the diffusion is
dominated by a 1-kick correlation not present in other kick
systems. At longer times, the global diffusion terms, though
weak, accumulate and eventually become dominant.

The method of correlations provides a generic and accurate way of
interpreting experimental data for this system
\cite{PJones}. There is also a simple physical picture. For
particles subjected to kicks of form $K \sin x$, consecutive kicks
will be  out of phase and will hence cancel if $p_0\epsilon \simeq
(2n+1)\pi$ where $n=0,1,2 \dots$. In other words, an impulse
$V'(x_j)=K \sin x_j$ will be immediately followed by another which
cancels it, since $V'(x_{j+1})=K \sin x_{j+1}\simeq K \sin (\pi
+x_j)$. This cancellation means that particles become trapped at
these momenta. In contrast, particles for which $p_0\epsilon
\simeq 2n\pi$, will experience enhanced diffusion.

It was shown in \cite{PJones} that the new types of global families
of correlations control the escape from, and through, these ``trapping 
regions''. An unexpected feature of the classical calculations (and 
also seen in experiment)
was the observation that particles initially prepared in the trapping regions
will eventually gain more energy than those initially prepared in regions of
enhanced diffusion, after a timescale $t \gg  1/(K\epsilon)^2$\cite{PJones}.

\section{Quantum dynamics and Floquet states}
If a Hamiltonian has a $T$-periodic time dependence, $H(t + nT) = H(t)$,
then the Floquet theorem  implies that 
solutions to the time-dependent Schr\"odinger equation
can be written in the form
\begin{equation}
\psi(t) = \exp\left[-i \epsilon t/ \hbar \right] \phi(t)
\end{equation}
where $\phi(t)$ is a $T$-periodic function called a Floquet state,
and $\epsilon$ is a quantity with dimensions of energy, termed
a quasienergy. This type of relation is familiar in the context 
of solid-state physics, where a Hamiltonian's invariance under 
discrete shifts of
{\em spatial} position (typically arising from a lattice structure)
allow solutions to be written analogously
in terms of Bloch states and quasimomenta.
The Floquet states provide a complete basis, and thus the time-evolution
of a general quantum state under periodic driving
can be expressed as
\begin{equation}
\Psi(t) = \sum_n  \  c_n \exp\left[-i \epsilon_n t / \hbar\right] \phi_n(t) ,
\label{Floquet}
\end{equation}
where $\{ c_n \}$ are {\em time-independent} expansion coefficients.
It is clear from this expression that the Floquet states and quasienergies 
play a similar role for periodically-driven systems to that
of energy eigenstates and eigenvalues in the time-independent case.

The time-evolution operator $U(t_2,t_1)$ may be used to
evolve a quantum state from time $t=t_1$ to a time $t=t_2$.
For a time-periodic system,
the single-period propagator $U(T,0)$ allows
a quantum system to be evolved ``stroboscopically'' at intervals separated
by the period $T$ with great efficiency, by defining the quantum
map $\Psi(n T) = \left[ U(T,0) \right]^n \Psi(0)$.
In terms of Floquet states, it is straightforward to show that
the time-evolution operator is given by
\begin{eqnarray}
U(t_2, t_1) &=& {\cal T} \exp\left[ -\frac{i}{\hbar} \int_{t_1}^{t2}
H(t') dt' \right] \label{define} \\
{ } &=& \sum_n e^{-i \epsilon_n t / \hbar}
| \phi_n (t_2) \rangle \langle \phi_n (t_1) | ,
\end{eqnarray}
where ${\cal T}$ is the time-ordering operator,
and thus it can be seen that 
the quasienergies and Floquet states can be conveniently
obtained by simply diagonalising the one-period propagator.
The eigenvectors of this operator are the
Floquet states, while its eigenvalues are related to the
quasienergies via $\lambda_n = \exp\left[-i \epsilon_n T / \hbar \right]$.

Obtaining an explicit form for this propagator
is normally a complicated procedure, as in general the
driving field does not commute with the static Hamiltonian.
For the case of $\delta$-kicking, 
however, the problem is simplified considerably which allows an analytic
form for the propagator to be written. For the QKP (Eq.\ref{ham_kp})
the propagator is given by
\begin{equation}
U(T,0) = \exp\left[ -i T p^2 / 2 \hbar \right]
\exp\left[ -i (K / \hbar) \sin x \right] .
\end{equation}
Using a basis of plane-wave states, 
the matrix-elements $\langle m+q | U(T,0) | n+q' \rangle$ 
of this operator can easily be shown to be 
\begin{equation}
U_{m,n}(T,0) = \exp\left[-i T (m+q)^2 / 2 \hbar \right] 
J_{m - n}(K / \hbar) \delta (q-q'),
\label{prop}
\end{equation} 
where $q$ is the quasimomentum ($p = (m + q) \hbar$) and
$J_n$ is the $n$-th Bessel function of the first kind.
For practical purposes it is useful to note that $|J_{m - n}(x)|$
decreases extremely rapidly with increasing $|m-n|$, thus giving
$U$ an effectively banded-structure. 

The single-period propagators for
the unequally-kicked systems can now be expressed as the product
of two matrices of this form, $U(T_{tot},0) = U(T_1+T_2,T_1) U(T_1,0)$,
where $T_1$ and $T_2$ are the two kick-periods. It should be noted
that although the single-kick propagators do not
conserve quasimomentum, their product does. As a consequence the evolution
of an ensemble of non-interacting particles which can be modelled by 
the evolution of a superposition of states with different quasimomenta,
represents a computationally efficient procedure: we can consider each 
quasi-momentum component independently. In our study of Floquet states
in effect this means we can diagonalise the smaller matrix given by \ref{prop},
which is block-diagonal in $q$.

Having obtained the Floquet states by diagonalising
(\ref{prop}), it is useful to analyse their structure,
particularly their  spread in momentum space.
We do this by evaluating a localization length, $L$ of each Floquet state 
at $t=0$.
We note that in this case we cannot assume that the Floquet states
have the usual exponential momentum distribution 
$N(p) \sim \exp \left[ -|p-p_0|/L \right]$,
of the usual $\delta$-KP. Hence we take  $L$ to be simply
the root mean square deviation from
the mean, $L=\sqrt{\overline{p^2}_n-{\overline{p}}^2_n}$.
In this expression
$\overline{p}_n$ is the mean momentum of the $n$-th Floquet state
at $t=0$,
$\overline{p}_n = \langle \phi_n(0) |p| \phi_n(0) \rangle$,
and $\overline{p^2}_n = \langle \phi_n(0) |p^2| \phi_n(0) \rangle$.
If the Floquet states do not have a strong time-dependence, such as
for the standard, or even the perturbed-period KP, this is adequate to
quantify the degree of spreading in momentum space. We shall
see, however, that for the case of the 2$\delta$ kicked
system it is not sufficient to measure the localization
at a single time, due to the extremely strong time-dependence
of the Floquet states.

\section{Results}

\subsection{Perturbed-period KP: a chaotic ratchet}

\begin{figure}[ht]
\includegraphics*[width=3.0in]{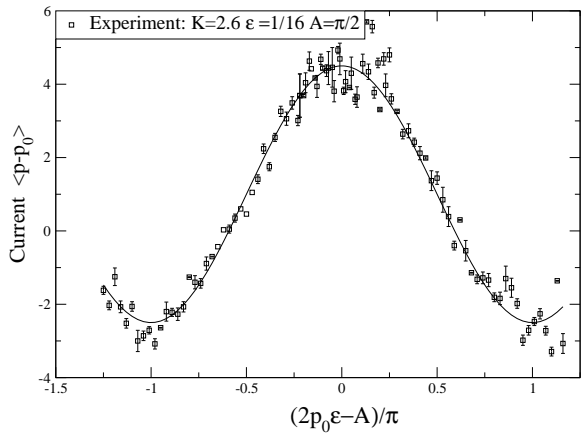}
\caption{Experimental values of the momentum current 
$I=\langle p-p_0 \rangle$,
for the perturbed-period KP, obtained with cold cesium atoms
in a pulsed optical lattice for $K \simeq 3$, $\epsilon=1/16$.
The solid line is a best-fit to the data,
showing that the current oscillates sinusoidally
as $I \propto \cos{2p_0\epsilon}$ for $A=\pi/2$.}
\label{Fig1}
\end{figure}

Fig.\ref{Fig1} shows a plot of the experimental ratchet current
obtained in \cite{Jon1}.  
A series of momentum
distributions $N(p-p_0)$ as a function of $p_0$ for a cloud of
cold cesium atoms in an optical lattice pulsed with unequal
periods.  The momentum distribution is essentially unchanged after 
about 60 kicks; the plotted values correspond to about $T=200$ kicks,
hence well after Dynamical Localization.
Full details are given in \cite{Jones,Jon1}, but by
employing an accelerated lattice, the experiment simulated an
effective rocking potential with $A \simeq \pi/2$. The first
moment of each localized distribution $I=\langle p-p_0 \rangle$ was then
calculated and plotted as a function of $p_0$.
In particular, a distribution centered at
$p_0=0$ initially, and with zero initial momentum current 
$\langle p \rangle  = 0 $
at $t=0$, yielded a finite and constant momentum current 
$\langle p \rangle \sim 4$ at long times. For non-zero
initial momenta $p_0 \neq 0$, an
oscillation $I \propto \cos(2p_0\epsilon)$, was observed and is
seen in Fig.\ref{Fig1}. This may be qualitatively understood from the form
of the classical two-kick momentum-diffusion
correction introduced in Sec.II, $ C_2= -K^2 \cos(2p_0\epsilon -\pi/2)$.
If we consider a very small momentum displacement $\delta p = p-p_0$,
differential absorption of energy for particles moving to the left $\delta p <0$
or right $\delta p > 0$ is proportional to the gradient
$\frac{\partial C_2}{\partial p_0} \propto K^2 \cos{(2p_0\epsilon)}$. An accurate
analytical form for the classical current was derived in \cite{Jonck}.

Of course, it follows that
a finite and persistent constant momentum current is also obtained classically.
It was found in \cite{Mont} that asymmetric diffusion persists only on a 
timescale $t \sim 1/(K \epsilon)^2$ and for this (unbounded) chaotic system, 
the acquired momentum asymmetry is never lost.
For a bounded (``compact phase-space'') system,
such asymmetries would vanish on a long time-scale,
since the distribution of a fully chaotic system would eventually 
become uniform. For this reason, until recently, it was argued that a fully 
chaotic system could not generate directed motion. 
So, although as shown in \cite{Mont}, the fully chaotic classical
system can keep a constant 
current for long times, practical implementation is ultimately less 
interesting since the 
average kinetic energy of the ensemble grows linearly with time and
without limit. Hence, this type of 
chaotic ratchet is of most interest as a {\em quantum} rather than a 
{\em classical} ratchet since in the quantum case DL
halts the diffusion and ``freezes-in'' the asymmetry, without the need for 
classical barriers like tori.

\begin{figure}[ht]
\includegraphics*[height=3.0in]{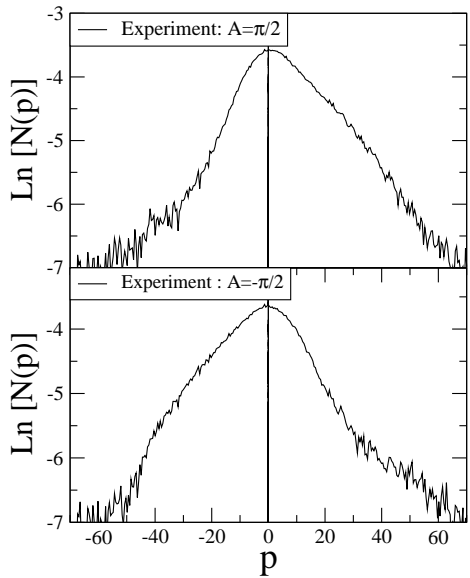}
\caption{Experimental momentum distributions $N(p)$ for the
perturbed-period KP obtained with cold cesium atoms in a pulsed
optical lattice for $K \simeq 3$, $\epsilon=1/16$ . The
distributions have localized, and hence remain essentially constant
with time. The results show clearly that the origin of the net
non-zero value of $\langle p \rangle$ obtained at long times is in 
the asymmetry of the DL profiles. As expected, the asymmetry is reversed 
by changing the sign of $A$, the amplitude of the rocking potential.}
\label{Fig2}
\end{figure}

In Fig.\ref{Fig2} we reproduce two
experimental momentum distributions for $K \simeq 3$
obtained with cesium atoms in Ref.\cite{Jones}, for $A=\pm \pi/2$. We clearly 
see that the origin of the non-zero momentum current is  the asymmetric 
momentum distribution.

As expected, Fig.\ref{Fig2} shows that changing the sign of
$A$ reverses the asymmetry. At this stage it may be unclear to the reader what the
significance of altering the sign of $A$ in the experiment might be, since 
after all, the rocking potential involves alternating impulses 
$K\sin x \pm A$. In fact the distinction (as may be ascertained
from the form of the classical diffusion) is between the case
where an impulse $K\sin x + A$ precedes free evolution for a
time interval $T_1 = 1 +\epsilon$ (obviously
followed by an impulse $K\sin x - A$ and interval $T_2 = 1-\epsilon$)
and the separate experimental case
where an impulse $K\sin x - A$ precedes free evolution for a time-interval 
$T_1 = 1 +\epsilon$ and so forth (which corresponds to a reversed current).

Note that the experimental range of $K \simeq 2.6 - 3.4$ does
correspond to a classical surface of section with some islands.
However, we note that classical quantities such as the average
energy are very accurately given by diffusion rates (with 2 and
3-kick corrections). The essential mechanism is asymmetric chaotic
diffusion: similar behavior was found at larger $K$ in \cite{Mont}
in regimes where there are no visible classical islands (but for
which experiments are not available); hence, in the analysis  of this 
type of ratchet, the presence
(or otherwise) of small stable islands is immaterial.
What is important, though, is that since the asymmetric diffusion
term is $2J_2(K)\cos(2p_0\epsilon-A)$, we need  $J_2(K) \neq 0$. Thus the 
much-studied (for the standard map) parameter value $K=5$ does {\em not} produce 
asymmetry, since $J_2(5) \simeq 0$. Values of $K \simeq 2.5 - 3.5$, 
$\hbar=1/4 - 1$,  on the other hand, turned out to be experimentally 
convenient and produced the strongest asymmetries.

\begin{figure}[ht]
\includegraphics*[width=3.0in, height=3.0in]{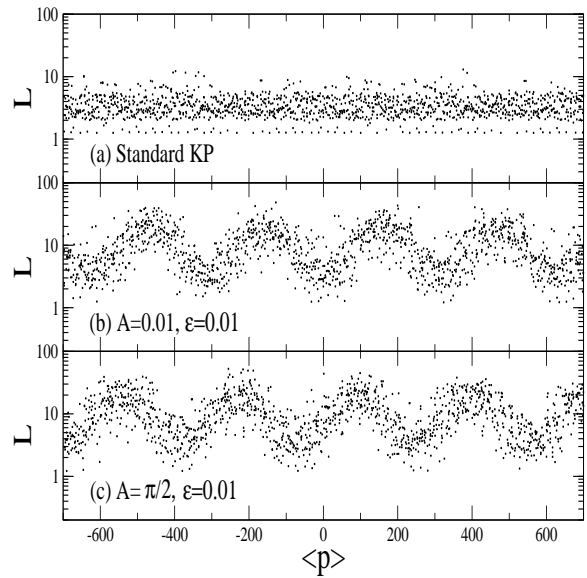}
\caption[]{The graph shows the localization lengths
$L$ of Floquet states as a
function of average momentum $\overline{p}$ (i.e. $\langle p \rangle$) for 
$K=3.4$, $\hbar$=1. Results are shown (a) for the standard QKP case, i.e.
$A=0.01$ (a non-zero $A$ was used to break spatial symmetry),
$\epsilon=0$, (b) for $\epsilon=0.01$, $A=0.01$ and (c)
$\epsilon=0.01$, $A=\pi/2$. The graph shows that for the standard
kicked rotor the $L$ are distributed within a narrow range in
comparison with other two below. For the rocking case, $L$
oscillates with $\overline{p}$ as expected from the 2-kick
correction $2J_2(K) \cos(2p_0\epsilon-A)$; the oscillations of the
two lower graphs are shifted relative to each other by a phase
$\pi/2$. The density of eigenstates corresponding to average
momentum range is roughly the same in all three cases.}
\label{Fig3}
\end{figure}

We now examine the form of the underlying Floquet states. In
Fig.\ref{Fig3} we compare the localization lengths for the
standard QKP, with those of the perturbed-period KP for
$K=3.4$, $\epsilon=0.01$.
The difference is quite striking; while
the standard QKP eigenstates are quite uniform across all regions of 
phase-space, the perturbed-period localization lengths oscillate 
sinusoidally with $\overline{p}$, with
a period of $\pi/\epsilon$. Introducing the additional rocking
potential with the accelerated lattice ($A=\pi/2$) clearly leads
to a $\pi/2$ shift in the oscillations. Inspecting
Fig.\ref{Fig3}(c) for  $\overline{p} \simeq 0$, we see that for
positive momenta the localization lengths are increasing, while
for negative momenta, the localization lengths decrease.
Note the nearly regular row of states for the standard QKP case
with $L \simeq 1$. These correspond to states localized on a
series of stable islands separated by $2\pi$, due to the momentum
periodicity of phase-space in that case.

We have chosen a parameter range for which $L \ll \pi/\epsilon$:
that is, the localization length of each state is much smaller than the
oscillation in $\overline{p}$. Hence individual Floquet states
really do sample ``local'' diffusion rates. We found that
if we move towards a regime where $L \sim \pi/\epsilon$, the
conclusions remain valid, but the amplitude of the oscillations is
considerably damped. Similarly, if the sign of $J_2(K)$ changes,
so does the sign of the sinusoidal oscillation.

\begin{figure}[htb]
\includegraphics*[width=3.0in,height=1.5in]{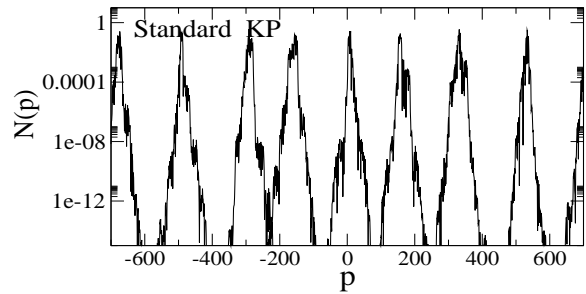} \caption{
Floquet states for the standard QKP, for $K=3.4$, $\hbar=1$. As
expected, all the states are exponentially localized, giving the
characteristic triangular shape of $N(p)$ when
plotted on a logarithmic scale.
They all have approximately similar localization lengths.}
\label{Fig4}
\end{figure}

We now consider the shape of the Floquet states in detail.
In Fig.\ref{Fig4}  we show the momentum distributions $N(p)= |\phi_n(p)|^2$
for Floquet states of the standard QKP.
The distributions (with $N(p)$ on a logarithmic scale) all show
the well-known triangular form \cite{Fish} -- the hallmark of dynamical
localization. It may be clearly seen that the localization lengths
vary little from state to state.

\begin{figure}[htb]
\includegraphics*[width=3.0in]{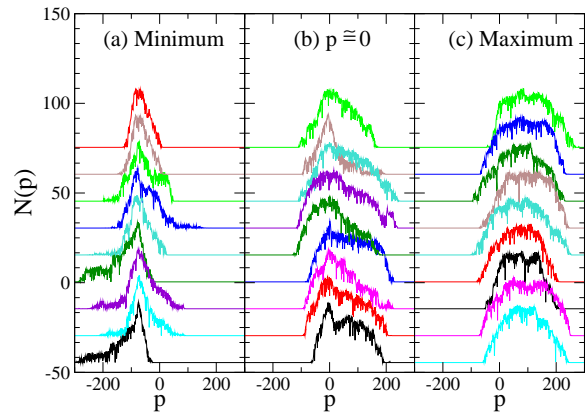}\\
\caption{Typical form of Floquet states for the perturbed-period KP,
$K=3.4$, $\epsilon=0.01$, $A=\pi/2$ and $\hbar$=1.
Here we plot $N(p)= |\phi_n(p)|^2$ as a function of $p$.
(a) states with $p \simeq -78$. This corresponds to a
minimum of the 2-kick correction $\cos{2p_0\epsilon -\pi/2}$.
The states are narrow, but in general, roughly symmetric.
(b) states with $p \simeq 0$. The typical state here is asymmetric
(c) states with $p \simeq +78$. This corresponds to a
maximum of the 2-kick correction. States here are generally symmetrical,
but broad and flat-topped.}
\label{Fig5}
\end{figure}

In Fig.\ref{Fig5}, by contrast, 
the localization lengths of the Floquet states of the perturbed period
$\delta$-KP display a strong dependence on the mean momentum of the states.
In addition, the figure shows
that states localized close to $p=0$ are markedly asymmetric.
The states are considerably extended towards positive
momentum, but are strongly localized towards negative $p$. This
behavior neatly accounts for
the form of the experimental momentum distribution
shown in Fig.1, which
for $A=\pi/2$ were also more extended towards positive $p$. The
states localized near $p \simeq  \pi/4\epsilon$  and
$\pi/4\epsilon$  correspond to, respectively, minima and maxima of
the classical diffusion. They are roughly symmetrical (typically)
but vary by up to a factor of $\sim 40$ in $L$. In Ref.\cite{Jonck}
it was proposed that the observed variation in the energy
absorption rates between atoms prepared with an initial drift
momentum $p_0= -\pi/4\epsilon$ (which absorb very little energy)
and those with $p_0= \pi/4\epsilon$ might be exploited to filter
traffic of atoms through an optical lattice. The form of the
underlying Floquet states explains this differential rate of
energy absorption.

Subsequently, it was found experimentally that the double $\delta$-KP 
in fact shows much more pronounced differential absorption rates, 
without requiring the application of a rocking field $A$.
We next report a study of the
Floquet states of this system.

\subsection{Double $\delta$-KP}

\begin{figure}[ht]
\includegraphics*[width=3.0in]{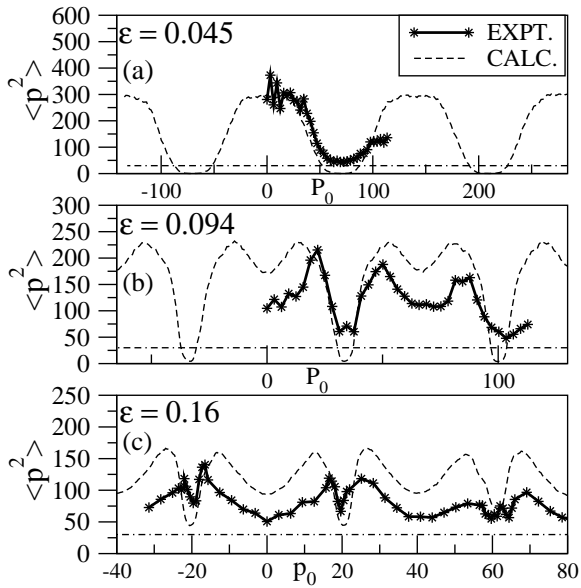} \\
\caption{Experimental results for double $\delta$-KP realization with
cesium atoms (see \cite{PJones} for details).
Each data point (star) shows the energy absorbed (after 100 kicks, 
$K=3.3$, $\hbar=1$)
by a cloud of atoms with average momentum $p=p_0$ (relative to the
optical lattice) at initial time, $t=0$. With increasing
$\epsilon$, we see the minima (maxima) in the energy flip into
maxima (minima) as a long-ranged family of classical correlations
gradually overtakes the 1-kick classical correlation. The dashed
lines represent a classical simulation using
100,000 particles, all with momenta $p_0$ at $t=0$, and $K$ within
the range $3.3 \pm 10 \%$. {\bf(a)} $t^* \ll t_1 \simeq 1/(K\epsilon)^2$.
Regime where a one-kick correlation is the dominant correction to the classical
diffusion. Here, atoms prepared near the trapping regions 
$(p_0 \epsilon \sim (2n+1)\pi)$ remain trapped.
Results follow closely the formula 
$\langle p^2  \rangle \simeq K^2t(1- \cos p_0\epsilon)$.
{\bf(b)}$t^* \sim 1/(K\epsilon)^2$. Regime showing the inverted peaks of
the Poisson correlation terms analyzed in \cite{PJones}, which determine the
momentum trapping very close to the resonant condition 
$(p_0 \epsilon = (2n+1)\pi)$.
{\bf(c)} $t^{\ast} > 1/(K\epsilon)^2$. Regime dominated by
correlation family $C_{G1}$, but sharp inverted peaks due to the Poisson 
correlations are still visible.}
\label{Fig6}
\end{figure}

A study of the experimental and classical behavior of the
double $\delta$-KP was carried out in \cite{PJones}. 
The classical
dynamics is very different to that of the perturbed period KP. At very
short times, the chaotic diffusion comprises an uncorrelated
diffusion term $K^2/2$ and one dominant 1-kick correction. It
was found in \cite{PJones} that one can approximate the growth in
the mean energy with time $t$, by the simple expression 
$\langle p^2 \rangle \simeq K^2t[1- \cos p_0\epsilon]$. 
In Fig.\ref{Fig6}(a) we show
experimental results for cesium atoms which localized in this
regime. The experiment measured the energy of a series
of clouds of $\sim 10^6$ atoms moving through the pulsed optical
lattice with varying average drift momenta $p_0$. For
Fig.\ref{Fig6}(a), the simple expression given above gives an
excellent fit to the experiment, if we take $t \sim t^{\ast}$, where
$t^{\ast}$ is the break time. This regime corresponds to 
$t^{\ast} \ll 1/(K\epsilon)^2$.

However, a more detailed study of the classical correlations showed
that for later times, a new type of correction appeared. Families
of long-ranged, or ``global'',
correlations which coupled {\em all} kicks appeared.
These corrections are individually very weak, but accumulate to
eventually dominate the diffusive process. One family (termed the
``Poisson family'' in \cite{PJones}) was shown to lead to 
well-localized, inverted peaks in the energy absorption at values of
$p_0 \simeq (2n+1)\pi/\epsilon$, where $n=0,1,2... \ $.
These values of $p_0$ correspond to trapping regions in phase-space 
(at low values of $K$, structures corresponding to islands
and broken phase-space barriers are evident). 
However there is no need to investigate
detailed transport through this complex mixed phase-space structure,
as the correlations
give us a generic and quantitative handle on the energy diffusion with time.
In this intermediate regime, dominated by the Poisson
correlations, atoms prepared
outside the trapping regions rapidly diffuse across the regions between them.
Particles prepared in the trapping regions remain there. This regime occurs for
$t^{\ast} \sim 1/(K\epsilon)^2$ and corresponds approximately to the experimental
results shown in Fig.\ref{Fig6}.

Finally, at the longest timescales, there is the $C_{G1}$
correction investigated in \cite{PJones}, which is a long-ranged
global-correlation family.
$C_{G1}$ results in an oscillation of the form $-\cos p_0\epsilon$ and becomes
dominant at the longest timescales. The oscillation is of the same
period as the 1-kick correlation, but is of opposite sign. This
means that at the longest timescales, the minima in energy
absorption shown in Fig.\ref{Fig6}(a) become maxima in energy
absorption; and vice-versa: the maxima become minima. Fig.6(c)
shows experiments tending towards this regime. The inverted peaks
of the Poisson family are still in evidence, but a $-\cos
p_0\epsilon$ oscillation is clearly superposed. This is a somewhat
counter-intuitive result since it implies that atoms initially
prepared in the momentum trapping regions are the ones which at
long times, for $t^{\ast} \gg 1/(K\epsilon)^2$, will absorb the most
energy (there are no further reversals of this behavior at even
longer times).

\begin{figure}[htb]
\includegraphics*[width=3.0in]{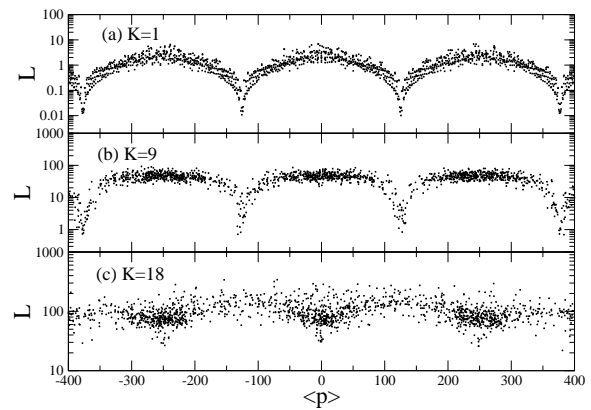}
\caption{Localization lengths of typical states for the
double $\delta$-KP ($\epsilon=0.025$, $\hbar=0.5$) corresponding to the three
classical diffusion regimes investigated in the experiments in
\cite{PJones}.}
\label{Fig7}
\end{figure}

The classical analysis thus reveals that there are
three distinct classical diffusive regimes
occurring at three timescales.
We can expect that the corresponding {\em quantum} behavior
will depend on which regime is dominant when dynamical localization
arrests the quantum momentum diffusion.
To investigate this, we now investigate how the form of the Floquet states
varies in these different regimes.

In Fig.7(a) we show the localization lengths of the Floquet states 
for a weak kicking-strength. It can clearly be seen that the localization
varies periodically as a function of momentum, staying within
the range $1 \leq L \leq 10$ for the majority of points, 
with the exception of a series of 
sharp cusp-like features at which the localization dramatically falls.
These location of these cusps exactly corresponds to the
the ``trapping momenta'', $p \epsilon = (2 n + 1) \pi$,
predicted from classical arguments.
The Floquet states centered in the trapping regions
have widths of $L\simeq 0.01$, much narrower than states
localized on stable islands, which are also visible in this figure
as regular strings of points at $L \sim 1$. 
At the experimental values of $K \simeq 3$ and $\epsilon =0.01$
a similar behavior is produced, with the broadest Floquet states
having localization lengths of $L \simeq 60$,
while the narrowest have widths of $L \simeq 0.03$, over one thousand
times narrower.

The Floquet states at the tips of the cusps have such low localization
lengths that they are effectively pure plane-wave states (this can be further
corroborated by evaluating the inverse participation ratio for these
states, which indeed takes a value of almost unity).
It is thus unsurprising that the presence of these states corresponds to
the classical trapping effect, as a quantum system prepared in
such a state will have a vanishingly small overlap with any other
state and so will remain frozen (or trapped) in its initial state.
It is important to note, however, that this quantum trapping effect 
depends critically on the {\em order} of the two kick- periods -- that 
is, whether the system is driven with a short-long
kick-sequence or the inverse long-short ordering.

\begin{figure}[ht]
\includegraphics*[width=3.0in]{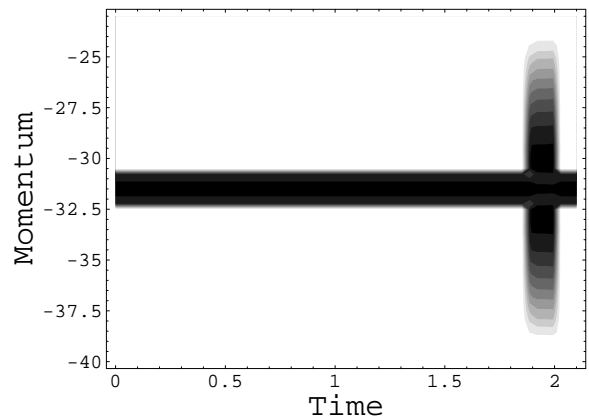}
\caption{Time evolution of a localized Floquet state
in momentum space, $N(p,t)$,
for physical parameters: $K=2$, $\epsilon=0.1$ and $\hbar=0.5$.
Initially the Floquet state is sharply peaked at $p = -10\pi$, in the
center of a trapping region. The first kick at $t=1.9$ causes the state
to spread across a much broader range of momentum, until the second kick
at $T=2$ restores the localized state.} 
\label{floq}
\end{figure}

\begin{figure}[ht]
\includegraphics*[width=3.0in]{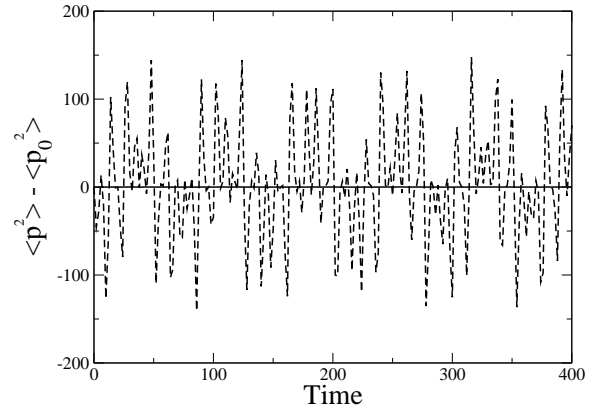}
\caption{Time-evolution of the energy of the double $\delta$-KP,
for the same physical parameters as in Fig.7a. The solid line shows
the evolution of the system under the long-short kicking sequence, and
shows little features. The dotted line shows the result of the short-long
sequence, and exhibits a complicated quasi-periodic behavior.}
\label{evolve}
\end{figure}

This may appear surprising at first, since the Floquet
states are periodic, with the same period $T_{tot}$
as the driving, and this period is not altered
by interchanging the order of the kicks.
Although it is frequently neglected, however, it is important to
recall that the Floquet states do have an explicit
time-dependence within each period,
and this {\em is} able to produce substantially different
behavior \cite{creffield} when the phase
of the driving field is altered. To illustrate this,
we show in Fig.\ref{floq} the
time-evolution of one of the localized Floquet states, which
experiences $\delta$-kicks at times
$t = T_1 = 1.90$ and $t = T_{tot} = 2$. As can be seen, the state
has only a trivial time-evolution during the
first time-interval ($0 \leq t < T_1$), since it is almost
a plane wave and is thus approximately an eigenstate of the free Hamiltonian.
The first kick at $T_1$ causes the wave-packet to spread considerably
in momentum space, before the second kick restores this broadened state
to its original narrow form.
Thus in this brief window of time between the two kicks, even the
most localized Floquet states have a considerable spread in
momentum. As a consequence, if the phase of the kicking
field is shifted so that
the system experiences the short-long kick-sequence, none
of the Floquet states are sharply localized in momentum at $t=0$.
We show in Fig.\ref{evolve} the time-evolution
of the system's kinetic energy
when it is prepared in a momentum eigenstate in a
trapping region. For the long-short kick-sequence this state projects
onto essentially a single Floquet state at $t=0$, and
so its time evolution is trivial and its energy
remains constant. For the case of the
short-long kick-sequence, however, the initial state projects onto a
number of Floquet states (Eq.\ref{Floquet}), giving rise
to a complicated quasi-periodic behavior arising from beating
between the different quasienergies.

In Fig.7(b), we see the effect of increasing
the kick-strength. In this regime 
we find there is an almost constant localization length
for momenta in between the trapping regions, which are again
signalled by sharp cusp-like structures. This
indicates that the Floquet states are confined between the classical
broken phase barriers in the trapping regions. 
Early studies
indicate that the level statistics of the
corresponding quasienergies are not pure Poisson in
this regime, as would be the case for the standard QKP.

In Fig 7(c), we see an inversion of the broad momentum modulation in
Fig 7(a), similar to the reversal seen in the experiment. In this
regime, the eigenstates localized in the trapping regions near $p \simeq
(2n+1)\pi/\epsilon$ are typically {\em broader} than those localized in
between.

\section{Conclusions}
We have presented a study of the Floquet states of $\delta$-kicked
particles pulsed with unequal periods, and used them
to analyze experimental data on these systems. We conclude that
the chaotic ratchet effect proposed in \cite{Mont} and observed
experimentally in \cite{Jones} is associated with asymmetric
Floquet states localized around $p=0$. We conclude also that the
behavior of the localization lengths of the Floquet states for the
double $\delta$-kicked rotor broadly accompany the change over between
the three distinct classical diffusion regimes investigated
experimentally in \cite{PJones}.


\begin{thebibliography}{99}

\bibitem{Casati}  G.~Casati,  B.V. Chirikov, F.M.~Izraelev F.M., and
 J. Ford in ``Lecture notes in Physics'', Springer, Berlin {\bf 93 }, 334
(1979); {B.V.~Chirikov, Phys. Rep. {\bf 52}, 263 (1979).}

\bibitem{Fish} S.~Fishman, D.R.~Grempel, and  R.E.~Prange, 
Phys. Rev. Lett. {\bf 49}, 509 (1982).

\bibitem{Raizen} F.L.~Moore, J. C.~Robinson, C.F.~Bharucha,
Bala~Sundaram, and M.G.~Raizen Phys. Rev. Lett.\ {\bf 75}, 4598 (1995).

\bibitem{Raizcat} D. A. Steck, W. H. Oskay, M. G. Raizen, Phys. Rev. Lett.\ {\bf 88}, 120406 (2002)

\bibitem{Phillips} W. K. Hensinger , H. Haffer, A. Browaeys, {\it et al.}.
    ``Dynamical tunnelling of ultracold atoms''
    Nature {\bf 412}, 6842 (2001) 

\bibitem{Izrael}F. M. ~Izraelev, Phys. Rep. {\bf 196}, 299 (1990)

\bibitem{Mont} T.S.~Monteiro, P.A.~Dando, N.A.C.~Hutchings
and M.R.~Isherwood, Phys. Rev. Lett {\bf 89}, 194102 (2002).

\bibitem{Jonck} T.~Jonckheere, M.R.~Isherwood and T.S.~Monteiro,
Phys. Rev. Lett. {\bf 91}, 253003 (2003).

\bibitem{Jones}P.H.~Jones, M.~Goonasekera, H.E.~Saunders-Singer and D.~Meacher,
preprint arXiv:quant-phys/0309149

\bibitem{Jon1}P.H.Jones {\it et al.}, to be published.

\bibitem{PJones}P.H.~Jones, M.~Stocklin, G.~Hur, and T.S.~Monteiro,
Phys. Rev. Lett. in press (preprint arXiv:physics/0405046)

\bibitem{Ditt} T.~Dittrich, R.~Ketzmerick, M.-F.~Otto, and
H.~Schanz, Ann. Phys. (Leipzig) {\bf 9},1 (2000);
H.~Schanz, M.-F.~Otto, R.~Ketzmerick , and T.~Dittrich,
Phys. Rev. Lett. {\bf 87}, 070601 (2001).

\bibitem{Flach} S.~Flach, O.~Yevtushenko, Y.~Zolotaryuk, Phys. Rev. Lett.
{\bf 84}, 2358 (2000).

\bibitem{Reimann} P.~Reimann, Phys.Rep.{\bf 361},57 (2002).

\bibitem{Shep} D.\L.~Shepelyansky Phys. Rev. Lett. {\bf 56}, 577 (1986).

\bibitem{Raizen2} B.G.~Klappauf, W.H.~Oskay, D.A.~Steck, and M.G.~Raizen,
Phys. Rev. Lett. {\bf 81}, 1203 (1998).

\bibitem{Garreau} J.~Ringot, P.~Szriftgiser, J.C.~Garreau, and D.~Delande
Phys.\ Rev.\ Lett.\ {\bf 85}, 2741 (2000).


\bibitem{creffield}
{C.E.~Creffield, Europhys. Lett. {\bf 66}, 631 (2004).}

\end{thebibliography}
\end{document}